\documentclass[sigconf]{acmart}
\AtBeginDocument{%
  \providecommand\BibTeX{{%
    \normalfont B\kern-0.5em{\scshape i\kern-0.25em b}\kern-0.8em\TeX}}}

\copyrightyear{2023}
\acmYear{2023}
\setcopyright{acmlicensed}\acmConference[CHI '23]{Proceedings of the 2023 CHI Conference on Human Factors in Computing Systems}{April 23--28, 2023}{Hamburg, Germany}
\acmBooktitle{Proceedings of the 2023 CHI Conference on Human Factors in Computing Systems (CHI '23), April 23--28, 2023, Hamburg, Germany}
\acmPrice{15.00}
\acmDOI{10.1145/3544548.3581099}
\acmISBN{978-1-4503-9421-5/23/04}

 \newcommand{\xhdr}[1]{\vspace{1.7mm}\noindent{{\bf #1.}}}

\usepackage{xspace}
\newcommand{\mverselib}{\texttt{multiverse}\xspace}

\newcommand{\ourtool}{\textsc{Multiverse Debugger}\xspace}
\newcommand{\template}{\textit{template}\xspace}

\def\mincover{\textsc{decision cover}\xspace}
\def\errorLogAggregation{\textsc{error message aggregation}\xspace}
\def\diff{\textsc{universe-to-multiverse diff}\xspace}

\def\rqWorkflowsLong{What workflows do analysts gravitate towards?\xspace}
\def\rqChallengesLong{What challenges do analysts need to overcome when debugging multiverse analyses?\xspace}
\def\rqToolAddressChallengesLong{To what extent \rr{do features like those in} \ourtool~address debugging challenges?\xspace}
\def\rqToolLong{How does \ourtool affect analysts' workflows?\xspace}

\newcommand{\shortquote}[1]{``\emph{#1}''}
\newcommand{\longquote}[1]{\vspace{-1pt}\begin{quote}``\emph{#1}''\end{quote}}

\newcommand{\rr}[1]{\color{black}{#1}\xspace\color{black}}

\usepackage[ruled]{algorithm2e}

\begin{document}

\title{Understanding and Supporting Debugging Workflows in Multiverse Analysis}


\author{Ken Gu}
\email{kenqgu@cs.washington.edu}
\affiliation{%
  \institution{University of Washington}
  \city{Seattle}
  \state{Washington}
  \country{USA}
}
\orcid{0000-0002-4343-1578}

\author{Eunice Jun}
\email{emjun@cs.washington.edu}
\affiliation{%
  \institution{University of Washington}
  \city{Seattle}
  \state{Washington}
  \country{USA}
}
\orcid{0000-0002-4050-4284}

\author{Tim Althoff}
\email{althoff@cs.washington.edu}
\affiliation{%
  \institution{University of Washington}
  \city{Seattle}
\state{Washington}
  \country{USA}
}
\orcid{0000-0003-4793-2289}




\begin{abstract}
Multiverse analysis—a paradigm for statistical analysis that considers all combinations of reasonable analysis choices in parallel—promises to improve transparency and reproducibility. Although recent tools help analysts specify multiverse analyses, they remain difficult to use in practice. In this work, we identify debugging as a key barrier due to the latency from running analyses to detecting bugs and the scale of metadata processing needed to diagnose a bug. To address these challenges, we prototype a command-line interface tool, \ourtool, which helps diagnose bugs in the multiverse and propagate fixes. In a qualitative lab study (n=13), we use \ourtool as a probe to develop a model of debugging workflows and identify specific challenges, including difficulty in understanding the multiverse's composition. We conclude with design implications for future multiverse analysis authoring systems.
\end{abstract}



\begin{teaserfigure}
  \includegraphics[width=\textwidth]{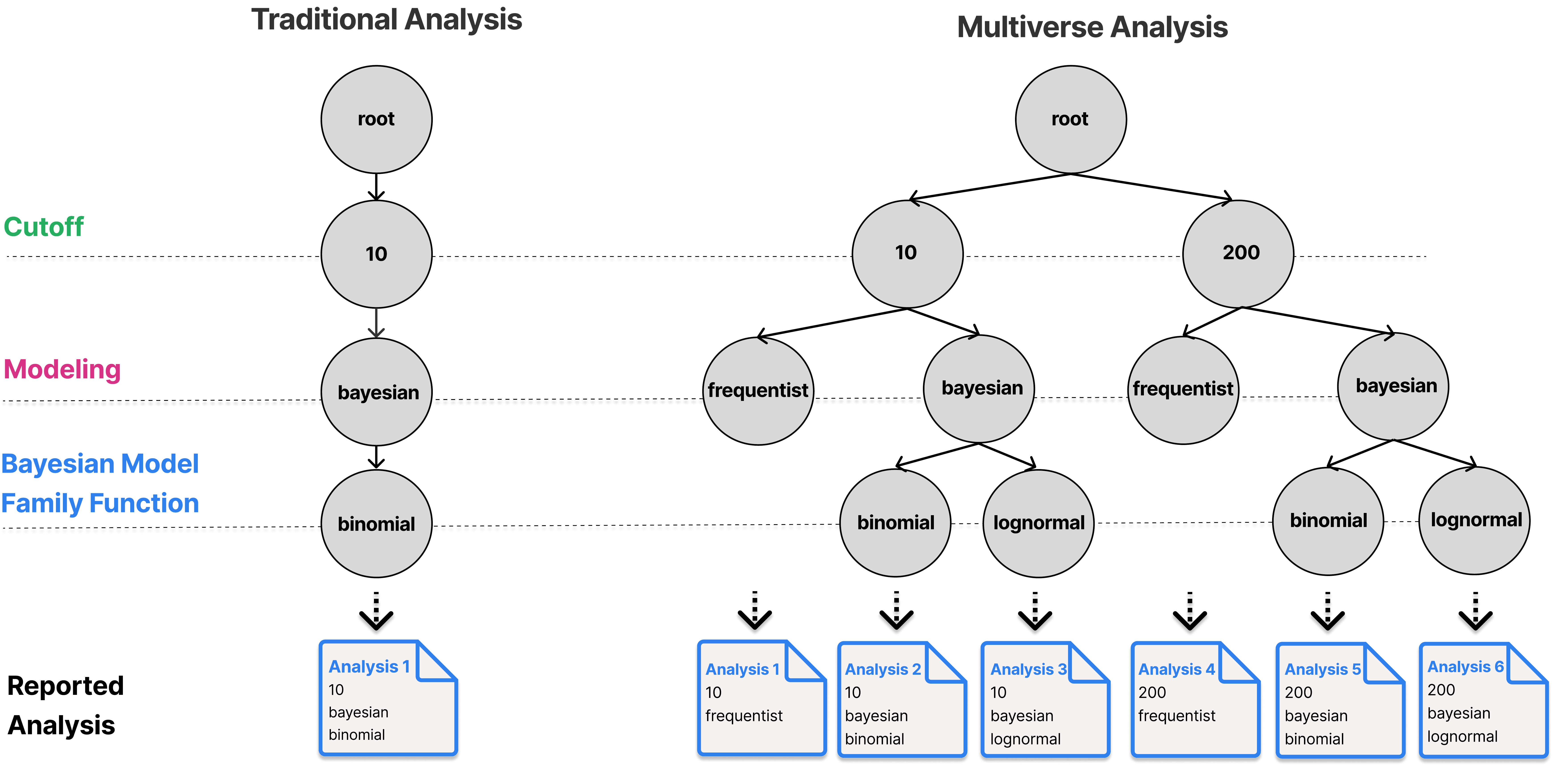}
  \caption{\textbf{Overview of Multiverse Analysis.}  In traditional analyses, an analyst may consider multiple decisions in their analysis---data filter cutoff, statistical modeling approach (e.g., frequentist, Bayesian), and Bayesian family function (e.g., binomial, lognormal). Traditionally, analysts may conduct multiple analyses with different decision choices but ultimately report only one combination of decisions (a “universe”). In contrast, in multiverse analyses, analysts consider, conduct, and report all reasonable combinations of decisions.
    }
  \Description{Multiverse Diagram}
  \label{fig:multiverse}
\end{teaserfigure}

\maketitle

\section{Introduction}
Even when trained analysts are given the same analysis task and dataset, they make different, sometimes conflicting, conclusions \cite{schweinsberg_same_2021, silberzahn_many_2018, patel_assessment_2015, breznau_observing_2022}. While it is not expected that different analysts when given only a dataset and a broadly defined task are to arrive at the exact same results, the level of variability is surprising. These divergences may even contribute to reproducibility crises across scientific disciplines~\cite{open_science_collaboration_estimating_2015, baker_1500_2016}. How could this be? Researchers believe that the flexibility in analytical choices (e.g., data filtering, statistical modeling approach, model parameters) is a key contributor. For example, analysts leverage their unique belief systems, domain knowledge, expertise, understanding of the problem, and exploratory results to justify their analytical decisions \cite{liu_paths_2020, kale_decision-making_2019}. Additionally, analysts only report the result of one set of analysis decisions despite exploring multiple combinations.

As a response to these problems, prior work has proposed multiverse analysis~\cite{steegen_increasing_2016, simonsohn_specification_2019} as a promising solution. Multiverse analysis, in contrast to traditional analysis,
is a statistical analysis paradigm that involves considering, specifying, and 
reporting results from all combinations of key decision options (\autoref{fig:multiverse} right). Multiverse analysis reveals how sometimes arbitrary decisions may affect an analysis conclusion. Moreover, by documenting and accounting for all reasonable decision options, multiverse
analysis, and related approaches such as sensitivity analysis, improve transparency and
robustness of statistical analyses and could prevent future reproducibility crises.

Despite the many benefits of multiverse analysis, authoring a multiverse
analysis remains challenging. Authoring multiverses is difficult because
analysts must explicitly enumerate decisions and the options for those decisions, write programs that generate
additional programs or scripts for each individual combination of options, compare and jointly interpret statistical results across all
combinations of decision choices, and iteratively debug and refine all the
above. Recent work in the HCI community and beyond provide tools to ease some of the challenges in the
authoring process:  Boba~\cite{liu_boba_2021},
\mverselib~\cite{sarma_multiverse_2021}, rdfanalysis~\cite{rdfanalysis}. However,
multiverse analysis remains difficult to adopt for many analysts. What are
authoring challenges that, if addressed, could lower the barriers to
authoring multiverse analyses? Prior work \cite{sarma_multiverse_2021} and our own correspondences with multiverse tool developers and multiverse practitioners have identified debugging as a central challenge.

In this work, we target multiverse debugging as a key challenge. 
Based on prior work~\cite{sarma_multiverse_2021}, our experiences, and with correspondences with multiverse practitioners and tool developers, we develop an initial model of debugging workflows in multiverse analysis (\autoref{fig:debug_workflow}). We find that analysts tend to focus on debugging a single analysis at a time (a ``universe''). Even debugging a single universe script is time-consuming due to the need to triage and fix code.
The scale of multiverse analyses, which can be on the order of tens of thousands of universes~\cite{liu_paths_2020}, exacerbates this problem and introduces additional cognitive burdens, such as
keeping track of how many unique errors there are, which set of universes these correspond to, and what portion of analyses
are buggy. Based on our initial workflow model, we identify three unique challenges of debugging in the multiverse paradigm: 
\begin{description}
\item[Challenge 1] --- Detecting bugs takes a long time during the slow execution of a multiverse (\autoref{fig:debug_workflow}D1).
\item[Challenge 2] --- Diagnosing the source of a bug to a specific decision choice or
set of choices (i.e., singular universe) is hard amongst thousands of universes (\autoref{fig:debug_workflow}D2).
\item[Challenge 3] --- After fixing a bug in a single universe (\autoref{fig:debug_workflow}D3), the analyst needs to remember changes and understand how to propagate them to the rest of the multiverse (\autoref{fig:debug_workflow}D4), which increases cognitive load and creates opportunities for error.
\end{description}
Although existing debugging tools and workflows help analysts fix a bug in a
specific universe, determining what universe to focus on and subsequently propagating one universe's changes to other universes that share the same error, remain under-supported.

To address these initial challenges, we prototype a debugging tool,
\ourtool(\autoref{sec:tool}). \ourtool is
a command-line interface (CLI) tool that extends Boba~\cite{liu_boba_2021}, an
an existing open-source tool that has already been employed in a large real-world study \cite{schuster_programming_2021}. \ourtool has
three key features, each of which addresses a challenge: (i) execution of a
a significantly smaller set of decision choice combinations to facilitate fast iteration (Challenge 1) (ii) aggregation of error messages across a multiverse analysis (Challenge 2), and
(iii) propagation of edits made to the rest of the multiverse (Challenge 3).

Using this tool as a probe, we conduct a qualitative lab study with 13 analysts to explore multiverse debugging in greater depth (\autoref{sec:focused_study}).
This lab study confirms Challenge 1 and Challenge 2 and we find \ourtool's features benefit analysts in diagnosing multiverse error messages and quickly detecting bugs.
We observe that Challenge 3 is not a central concern to analysts as, prior to propagating bug fixes, 
analysts already struggle with understanding the composition of the multiverse (i.e., the multiverse analysis tree in \autoref{fig:multiverse}), which is critical in their efforts to diagnose multiverse error messages. 
We also observe analysts, inspired by \ourtool, favor selective execution of a subset of universes in the debugging process, which current tools do not yet support. 


Based on these findings (\autoref{sec:label_results}), we update and extend our model of the multiverse debugging workflow and associated challenges (\autoref{fig:updated_workflow}). In addition, we discuss (\autoref{sec:discussion}) a set of design implications that include helping analysts better understand the composition of the multiverse and supporting analysts in navigating their multiverse analysis.

This paper contributes the following:
\begin{enumerate}
  \item Findings from a qualitative lab study that reveal
  open challenges in multiverse debugging,
  \item A publicly available open-source prototype of \ourtool that addresses some of these challenges and
  lab study results that evaluate to what degree our prototype's features can alleviate them~\footnote{The code
  for our prototype is publicly available at  \url{https://github.com/behavioral-data/multiverse-tooling}.}, 
  \item A model of the key operational steps in multiverse debugging workflows and associated challenges, and
  \item A set of design implications for how to better support debugging for multiverse analysis authoring.
\end{enumerate}

\section{Background and Related Work}

\subsection{Debugging in Software Engineering}
Debugging is challenging and time-consuming. In prior works aimed to understand debugging in software engineering, developers reported spending 20\% to 60\% of their time debugging \cite{beller_dichotomy_2018}. This has been later confirmed in a study analyzing real-world developer debugging sessions \cite{alaboudi_exploratory_2021}. 

A central challenge in debugging is the "large temporal or spatial chasms between the root cause and the symptom" \cite{kleiman_tales_1993}. Based on a prior lab study, researchers detailed the mechanisms of debugging as involving the processes of searching, relating, and collecting information of perceived relevance, in which the development environment plays a central role in influencing developers’ perceptions \cite{ko_exploratory_2006}. In other studies on general software development, it was discovered that a significant amount of mental effort is spent in understanding how a program works via searching relevant software artifacts, and inspecting source code/documentation \cite{minelli_i_2015, xia_measuring_2018}. 

With this understanding, multiverse debugging is likely to exacerbate the problems of traditional debugging workflows. There are more analyses to work with, more meta-data per analysis in the form of associated decision options which can affect the presence of bugs, and shared relationships between the collection of scripts that need to be considered. All this information, if not presented well, can make the process of collecting and relating relevant information significantly harder. We contribute the first user study to explore and model debugging behavior and challenges in the context of a multiverse analysis workflow.

\subsection{Multiverse Analysis}
Multiverse analysis \cite{steegen_increasing_2016, simonsohn_specification_2019} aims to have the analyst consider all reasonable decisions and combinations of decision options a-priori while then conducting and reporting all considered analyses. "Reasonable" here means actions with firm theoretical and statistical support \cite{simonsohn_specification_2020}.
Moreover, a decision in the multiverse paradigm is any decision an analyst may consider in an analysis. These decisions (e.g., Cutoff, Modelling, and Bayesian Model Family Function in \autoref{fig:multiverse}) are wide-ranging and can cover data collection and wrangling, statistical modeling, inference, and evaluation. For each decision, there are decisions options, defined as the alternatives that the specific decision could take. 

Because multiverse analysis considers all reasonable combinations of decision options, there is a combinatorial explosion in the number of universes as more decisions are involved. For example, a multiverse of 5 decisions each with 4 options would result in \begin{math}4^5 = 1024\end{math} universes. Prior work  has estimated that multiverses in practice contain between hundreds and hundreds of thousands of individual analyses \cite{liu_paths_2020}. 

As multiverse analysis has gained recognition and adoption \cite{rae_predicting_2019, poarch_effect_2019, kalokerinos_differentiate_2019, cesario_is_2019,  border_no_2019, dejonckheere_bipolarity_2018}, associated workflows, tools, and visualization techniques have been developed \cite{liu_boba_2021, sarma_multiverse_2021, rdfanalysis, dragicevic_increasing_2019, hall_survey_2022}. 
Recent work on multiverse authoring has identified debugging as an important, unaddressed challenge~\cite{sarma_multiverse_2021}. 
The present work extends this prior work by contributing the first user study and first prototype specifically focused on the unique debugging challenges that the multiverse paradigm presents.

\begin{figure*}[h]
  \centering
  \includegraphics[width=\linewidth]{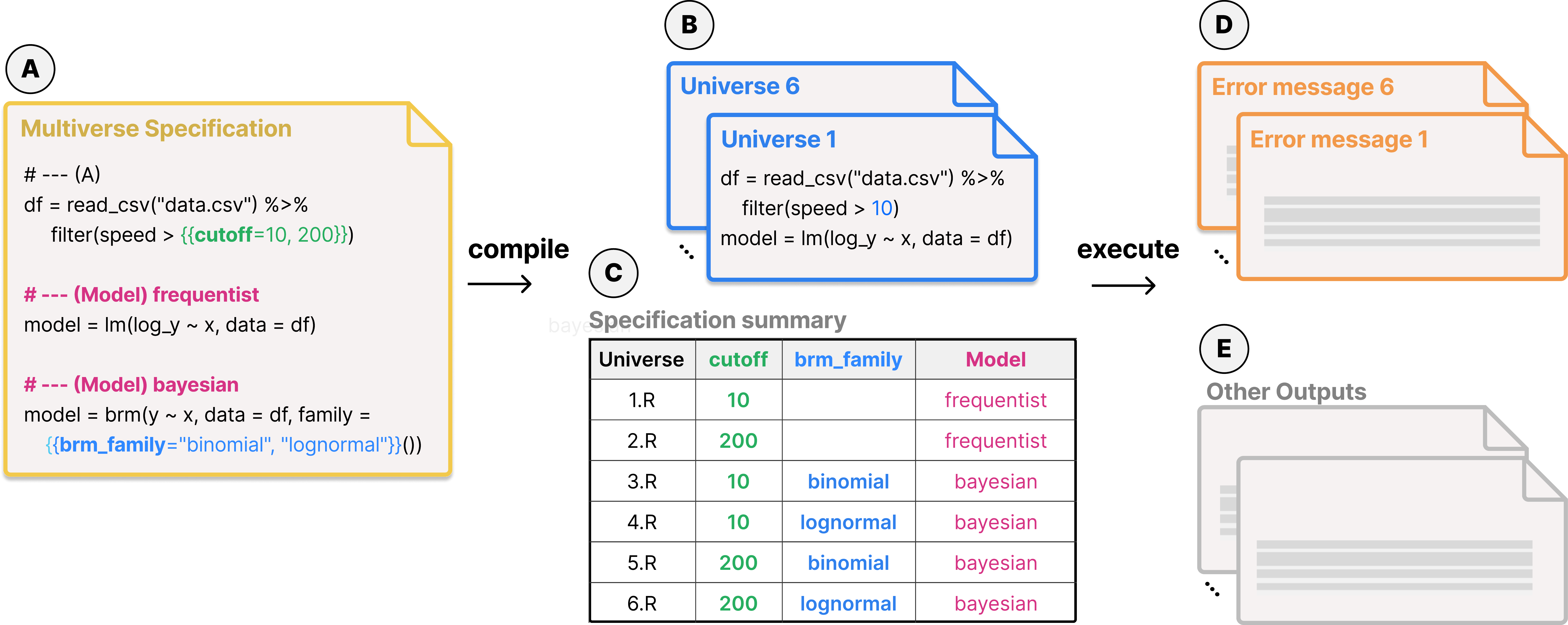}
  \caption{\textbf{Multiverse Authoring Process} An analyst starts out by writing a multiverse specification (A). Afterwards, the analyst compiles the multiverse specification into individual universes (B), which are enumerated in a specification summary (C). The specification summary indicates what the decisions are for a given universe. Next, when the universes are executed, each universe generates an error message (D) and other outputs (E) such as a model fit summary, model predictions etc. While this example shows Boba's domain specific language, other tools follow a similar process.}
  \Description{Multiverse Authoring Workflow}
  \label{boba_authoring}
\end{figure*}

\subsection{Tools for Multiverse Analysis}
Traditionally, analysts must consider hundreds if not thousands of universes  if they were to perform a multiverse analysis. This can result in a large set of mostly similar universe scripts which, with so many variations, is difficult to maintain \cite{kery_variolite_2017}. On the other hand, an analyst can write a complex series of control flow logic in one large script \cite{vanpaemel_multiverse_2016} but this makes it hard to selectively run an individual universe. Multiverse authoring tools make it easier to specify a multiverse analysis and execute it. These tools simplify specifying decisions by introducing special syntax to specify decisions, decision options, and constraints between decision options in one central file. In these general multiverse authoring tools, namely, Boba \cite{liu_boba_2021} and \mverselib \cite{sarma_multiverse_2021}, a common authoring workflow is observed (\autoref{boba_authoring}). Analysts specify their multiverse in a central multiverse specification containing different code snippets for different decision options (\autoref{boba_authoring}A). Afterwards, the multiverse specification is compiled into universes (\autoref{boba_authoring}B). A universe contains an instantiation of decisions' options and compilation also produces a specification summary enumerating each universe's decision options (\autoref{boba_authoring}C). After the compilation step, the universes are executed which each produces an error message (\autoref{boba_authoring}D) and other outputs (\autoref{boba_authoring}E).

While largely following the authoring workflow in \autoref{boba_authoring},  \mverselib aims to support the iterative workflow of a computation notebook. It is a R package that works in RMarkdown notebooks. To specify decisions, execute the universes, and gather results, analysts call \mverselib methods. The notebook acts as the multiverse specification. The compilation is implicitly performed under the hood when universes are executed.

Meanwhile, in Boba, the multiverse specification is one central \template file. Boba places specific domain-specific language (DSL) directives that indicate how different chunks of code fit together. This has the benefit of being programming language agnostic, treating non-multiverse code as raw strings. Nevertheless, because of these directives, the \template file is not executable and cannot leverage any of the advanced debugging features in modern interactive development environments (IDEs). Boba provides additional command-line commands to compile and run the multiverse. Analysts run \texttt{boba compile} to compile their multiverse
specification. To execute the multiverse after compilation, Boba provides the command \texttt{boba run} to execute a range of or
all the universes. When executed with Boba, each universe’s standard output messages and standard error messages are saved to a corresponding output file and can be gathered in a CSV file.

However, other than collecting error messages as entries in a table, both tools do not provide any additional support for multiverse debugging workflows. Our work extends the authoring framework of existing tools to study and alleviate the challenges encountered during multiverse debugging. In this paper, we focus on studying debugging workflows with Boba, as it is widely researched in the research community \cite{schweinsberg_same_2021, bloom_into_2022-1, niso_open_2022, gadhave_predicting_2020, smith_full_2022}. 
One advantage of Boba is that it is programming language agnostic, allowing multiverse analysis to reach a greater audience. 

\subsection{Debugging is a Challenge in Multiverse Authoring}
\label{subsec:workflow_model_exploratory}

\begin{figure*}[t]
  \centering
  \includegraphics[width=\linewidth]{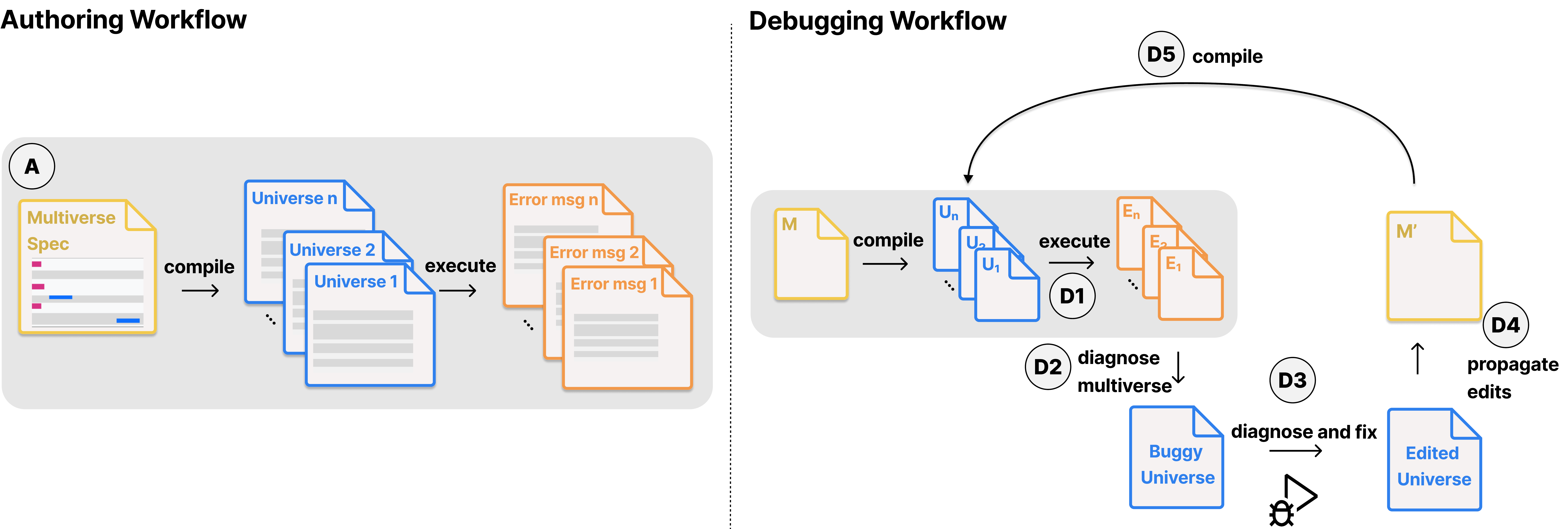}
  \caption{\textbf{Authoring vs Debugging Workflow}. Existing tools focus on the authoring process in a multiverse workflow (A) comprising of a multiverse specification, compiling the specification to universes, and executing the universes. However, there is an entire debugging workflow pertinent to the multiverse paradigm that is not well understood, presents challenges, and lack support from existing tools. \rr{We hypothesize} the following debugging workflow. First, an analyst executes universes which can generate error messages (D1). Here, detecting errors quickly is challenging because executing all universes is time consuming and a universe containing error prone code may not be executed until hundreds or thousands of others have been executed already. Next, the analyst diagnoses what decision or set of decisions caused the errors (D2) which guides them to focus on one buggy universe. This step is challenging because an analyst needs to synthesize information from a myriad of sources (i.e., the multiverse specification, universes, error messages, and the specification summary) which only gets worse as the multiverse scales. Now, debugging at the universe level, the analyst diagnoses and fixes their bug in the typical single script debugging paradigm and is free to use debugging tools they are most comfortable with (D3). Once the fixes are made at the universe level, the analyst then propagates the edits made to the universe back to the multiverse specification (D4). This step contains the challenge of remembering changes in the universe and where those changes propagate to in the overall multiverse specification. Finally, the analyst compiles the new specification (D5) and the cycle repeats. The gray area highlights shared workflow steps.}
  \Description{Initial understanding of workflow}
  \label{fig:debug_workflow}
\end{figure*}

Based on prior work~\cite{liu_boba_2021, dragicevic_increasing_2019, liu_paths_2020, sarma_multiverse_2021}, our experiences, and initial correspondences with multiverse practitioners and multiverse tool developers (see \autoref{appendix:formative}), we hypothesize an initial debugging workflow (\autoref{fig:debug_workflow}).
%
%
The workflow model is a first attempt to understand debugging in multiverse analysis and contrasts with the multiverse authoring workflow that is currently supported through existing tools (\autoref{fig:debug_workflow}A). After specifying and compiling a multiverse specification, the analyst executes the universes which produce error messages (\autoref{fig:debug_workflow}D1). Next, the analyst tries to diagnose the cause of the bug, which leads them to a single buggy universe to target (\autoref{fig:debug_workflow}D2). This step often involves examining multiple universes that share an error message. Once the analyst is working with an individual universe, they address the bug and make edits along the way, the same as they would when debugging a single script (\autoref{fig:debug_workflow}D3). After, they abstract and propagate the specific changes in the universe back to the higher-level multiverse specification (\autoref{fig:debug_workflow}D4). Finally, after the edits are propagated to form a new multiverse specification, it is compiled (\autoref{fig:debug_workflow}D5). This iterative debugging cycle typically repeats multiple times until all bugs are addressed.


This workflow model suggests the following three challenges to debugging a multiverse analysis.
 
\xhdr{Challenge 1 - Detecting bugs is slow} During the execution of the multiverse (\autoref{fig:debug_workflow}D1), the order of execution of the universes is arbitrary. Therefore, to discover a bug that occurs in a select few universes, hundreds or thousands of universes may need to be executed before the buggy universe is encountered. Even with running universes in parallel, this process can be time-consuming and drastically slows down the debugging cycle.

 \xhdr{Challenge 2 - Sifting through error messages and multiverse artifacts to diagnose a bug is difficult at scale} In the process of diagnosing an error from running the multiverse (\autoref{fig:debug_workflow}D2), an analyst potentially needs to navigate through many error messages, many universes, and the specification summary and relate these sources to understand the shared decision options of an error. It is infeasible for an analyst to inspect hundreds of files (or a single file that combines these) and looking at a significantly smaller subset may not fully isolate the shared decision options of an error and divert focus from the true source of a bug. We note that a multiverse that does not lead to any error messages is not necessarily bug-free. For example, a poorly specified model formula may not be statistically sound but may not raise any error messages. However, many bugs exhibit themselves as error messages and that is the primary way analysts debug in our experience.

 \xhdr{Challenge 3 - Abstracting and propagating universe edits to the multiverse increases cognitive load} In the procedure to abstract and propagate universe edits to the multiverse specification (\autoref{fig:debug_workflow}D4), the analyst needs to remember all their edits and locate where to place them in the multiverse specification. In complex multiverse specifications and universe edits that involve many changes, propagating universe edits induces additional cognitive demands, especially when the analyst must keep track of the associated decision options underlying the code they are propagating.

\section{Prototype: \ourtool}
\label{sec:tool}
To better understand multiverse debugging workflows and how to support them, we set out to build a prototype tool, \ourtool, to use as a probe in our subsequent lab study. We identify three design goals to support analysts in the multiverse debugging workflow (\autoref{fig:debug_workflow}). The goals correspond to addressing the three challenges identified in \autoref{subsec:workflow_model_exploratory}. 

\begin{enumerate}
    \item \textit{DG 1 - Reduce the time between executing universes and encountering error messages.}
    After compiling a multiverse specification, a tool should enable analysts to quickly observe error messages before committing to running the full multiverse. This is in the spirit of unit testing in which different components of the multiverse are rapidly tested before running the entire system. Quickly identifying error messages before executing an entire multiverse may reduce time spent authoring (buggy) multiverse analyses.
    
    \item \textit{DG 2 - Give an overview of error messages and how they relate to specific
    decision options.} Running thousands of universes can lead to thousands of individual error messages. In addition, error messages may arise due to a combination of decision options, which the analyst did not test when writing the multiverse specification. Therefore, diagnosing the severity and frequency of an error message helps to identify which parts of the code may need to be updated (including adding or removing dependencies between decision options in the multiverse specification). 
    To identify common bugs and distinguish among
    different kinds of bugs, summarizing the frequency of error messages and connecting
    them to specific decisions and decision options are likely to help
    analysts.
    \item \textit{DG 3 - Support the abstraction and propagation of single universe bug fixes to a multiverse specification.} The context of a
    multiverse analysis adds new complexity to fixing bugs. An analyst may elect to debug error messages in a specific universe as opposed to the higher-level multiverse specification. This enables the analyst to take advantage of already familiar and idiosyncratic ways of debugging specific universe error messages. In the analyst's process of debugging a single universe, they can leverage an entire ecosystem of single script debugging tools that they may already be familiar with. Therefore, making the process of propagating changes to individual universes to the higher-level multiverse specification easier, empowers the preferred single universe debugging workflow. 
    
\end{enumerate}
Based on these three design goals, we implement \ourtool with three core features: \mincover, \errorLogAggregation, and \diff.
The features
of \ourtool  are designed to be used after compiling a written multiverse specification. 
This prototype extends the Boba multiverse library \cite{liu_boba_2021} and each feature is exposed through the Boba command line interface.

While we implemented \ourtool on top of Boba, the challenges and design goals would largely exist for other multiverse authoring tools as well. Boba makes the decision to represent universes and error messages as individual files. While other tools may make different design decisions such as consolidating all these into a single file or object, this would still result in similar challenges of slow detection of bugs (Challenge 1) and difficulty of diagnosing error messages from a large number of universes (Challenge 2). These challenges are ubiquitous because of the combinatorial explosion of universes which is inherent in multiverse analysis' definition to run individual analyses corresponding to all combinations of decisions. Therefore, these challenges which motivate DG 1 and DG 2 persist no matter the choice to represent universes as individual files or some other format. Challenge 3 and DG 3, meanwhile, are more specific to a universe level workflow which is enabled by tools like Boba in which the universe is represented as a single file. However, the choice of whether a tool enables a universe level workflow or multiverse level workflow (in which individual universes are not easily editable) comes with its own trade-offs which we further describe in \autoref{sec:future_work}.

Both the \errorLogAggregation and the \diff interfaces are
implemented as web applications in Python. The frontend uses HTML, CSS and
Bootstrap \cite{bootstrap}, and the backend uses Flask~\cite{flask}. The \diff interface also uses the Monaco Editor library~\cite{monaco_editor}.

\subsection{Accelerating Bug Discovery Through Minimum Cover Approximation} 
\label{sec:mincover}
A key problem in executing universes with existing tools is the latency between executing universes and encountering error messages. Analysts may not encounter a universe that contains a specific decision option until hundreds or thousands of universes have already been run. \mincover can reduce the latency in detecting a bug (DG 1) by helping the analyst quickly identify all error messages corresponding to code in a specific decision option while running a much smaller subset of universes. In seven multiverses we tested, \mincover reduced the number of universes to run by over 98\%. After the analyst runs \mincover (\texttt{boba run ----cover}), \mincover calculates the reduced set of universes, executes them, and surfaces the \errorLogAggregation interface (\autoref{sec:erroraggr}) to summarize the error messages encountered in the executed universes. The analyst can interact with this interface to promptly see the set of error messages caused by a bug in any decision option.

\mincover calculates an approximation to the minimal set of universes to run such that all decision options in the multiverse are "covered". The problem of finding the minimal set of universes reduces to the classic set cover problem \cite{karp_reducibility_1972} which is known to be NP-hard \cite{approximation_2006}. To encourage trying different universes during each \mincover run, we employ a heuristic approximation based on random sampling that is highly effective in practice. We describe the \mincover algorithm in detail in \autoref{appendix:min_cover}. 

Making sure each decision option is encountered corresponds to ensuring ``condition coverage'' in traditional software testing \cite{test_oop}. However, \mincover does not ensure ``multiple-condition coverage'' \cite{test_oop} which would require running all combinations of decision options (essentially the entire multiverse) and leads to the combinatorial explosion of execution time. Nevertheless, error messages raised by "multiple-condition coverage" but not "condition coverage" are rare and become more obvious after the errors \mincover raises are addressed.

\begin{figure*}[t]
  \centering
  \includegraphics[width=\linewidth]{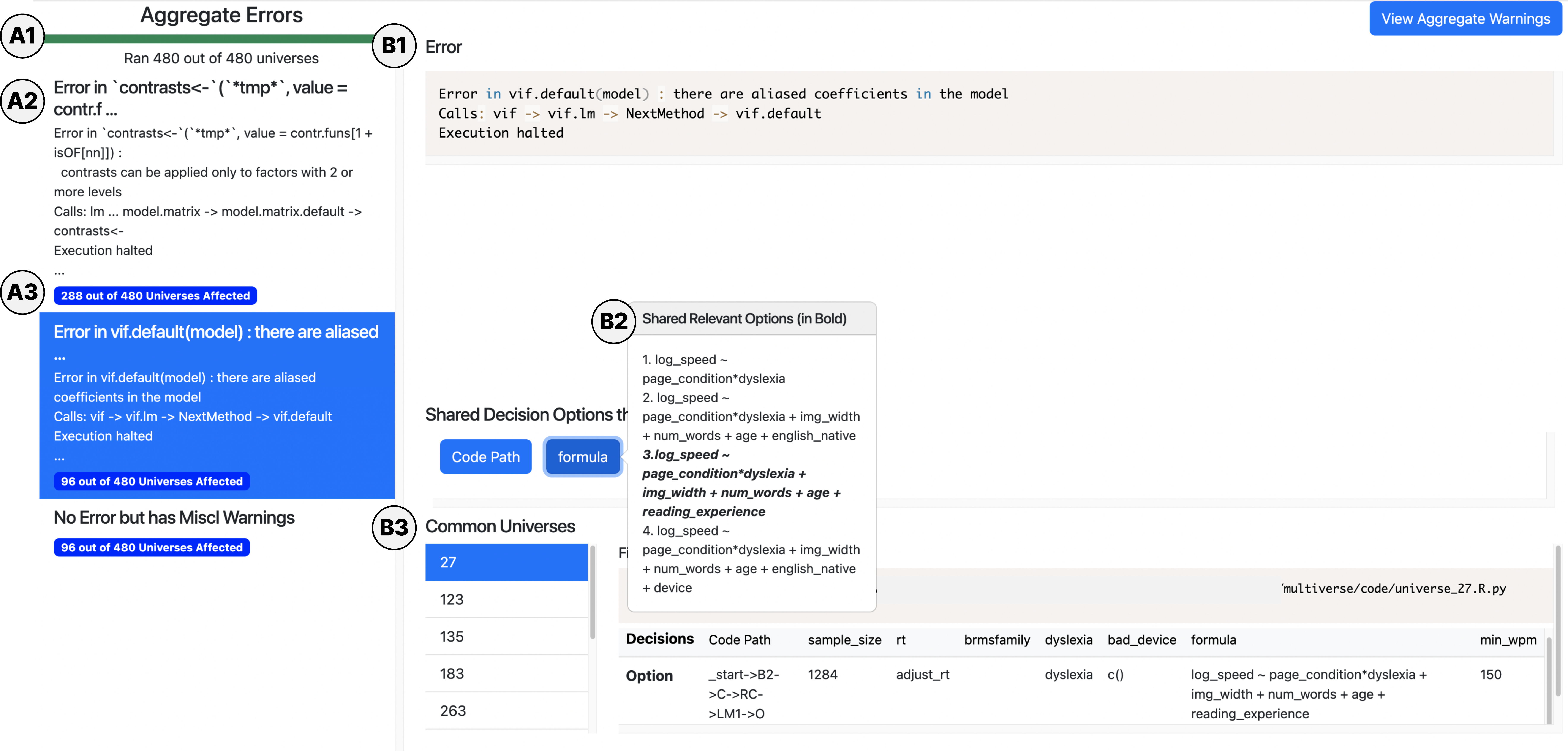}
  \caption{\textbf{Error Message Aggregation Interface}. The left side panel (A) contains a progress bar (A1) and unique groups of error messages in the universes ran so far. Each group contains a preview of the error message (A2), and the number of universes affected in each error aggregation (A3). The left side panel selects the error message to look into in the main panel (B). The main panel comprises of the full error message (B1), the decision options that are shared in the error aggregation (B2), and sample universes that contain the error message (B3). Without such a tool, analysts would have to manually inspect error messages (potentially hundreds or thousands of error messages) while cross-referencing universe entries in the specification summary. Not only is this action tedious but it is prone to missteps leading to poorly understood bugs. \errorLogAggregation seeks to address this challenge by automatically surfacing the information of all unique error messages and shared decisions in a grouped error.}
  \Description{\errorLogAggregation GUI}
  \label{fig:error_aggregate}
\end{figure*}

\subsection{Diagnosing Bugs via Error Message Aggregation}
\label{sec:erroraggr}
A core challenge in diagnosing a bug from error messages is that it is difficult to sift through the myriad of information sources (i.e., error messages, universe files, and the specification summary) to diagnose a bug to a set of decision options. Therefore, we design \errorLogAggregation to aggregate this information automatically and give analysts an overview of error messages and how they relate to specific decision options (DG 2). \errorLogAggregation supports two interactions: identifying groups of error messages and the scale of an error, and understanding the decisions that may cause an error. 

When an analyst runs the \errorLogAggregation command (\texttt{boba ----error}), the program ingests error messages from executed universes and categorizes errors based on string similarity (to handle slight line number or other changes in the error traceback). String similarity is calculated using the \texttt{string grouper} Python library \cite{berg_string_2022} and is based on the cosine similarity of vectors of TF-IDF values in which the terms are N-Grams. Afterwards, the information is presented in an interactive interface that includes the aforementioned interactions~(\autoref{fig:error_aggregate}). 

\subsubsection{Identifying error groups and the scale of errors}
The analyst can quickly identify
the number of universes affected by each error in a summary panel on the
left-hand side (see~\autoref{fig:error_aggregate}A1-3). Each error group has a preview of the error
text~(\autoref{fig:error_aggregate}A2) and a badge indicating the number of universes affected~(\autoref{fig:error_aggregate}A3). The
panel also displays a progress bar indicating the progress of universes run so
far and updates when the user refreshes the page ~(\autoref{fig:error_aggregate}A1). The summary panel gives the user a sense of how many errors occur relative to what universes have already been executed. In addition, the summary view of error groups helps the analyst assess a bug's frequency across the multiverse and subsequently prioritize errors.

\subsubsection{Understanding the decisions that may cause an error}
Once an analyst has selected an error to investigate from the summary panel, they can focus on the shared decision options that potentially
isolate an error group via the center panel (\autoref{fig:error_aggregate}B1-3). 

The center panel shows a traceback of the error message as well as the shared decisions options of all the universes that caused that error (\autoref{fig:error_aggregate}B2). Each decision that may cause the error is shown as a button to the analyst in which they can then click to see the shared decision options of universes that have this error message. Decisions that are most likely irrelevant to the error are removed to shift focus to the potential buggy decisions. 

To determine whether a decision is irrelevant the following heuristic is used.
If the error involves all options of a decision, then it is unlikely that anything in that decision is
causing the error. If the error involves not all options in a decision, then 
there is a possibility that something specific to that option could affect the
error. It must be noted, however, that the existing heuristic may not work if
each option has an error that is identical. 
However, this scenario may be unlikely and it was never encountered throughout our entire study.

With a better understanding
of the severity and decision options associated, the analyst can focus on
a specific universe that has the selected error message to fix the specific bug. With an
understanding of shared decision options in an error, the analyst may be able to better 
isolate where the error occurs and start with a more focused understanding of
how the error may have occurred. Moreover, having a grasp on the isolated decision
option that may cause an error provides a more semantically meaningful
error message than a single script bug. With the additional information of an
error message, the analyst can look for common universes that share the error at the
bottom of the main panel (\autoref{fig:error_aggregate}B3) and begin focusing on one universe. 
Overall, the emphasis on outlining shared decisions across an error message can potentially help the analyst focus on a specific universe and the code most likely to cause the error.

\subsection{Propagating Universe Edits with Universe-to-Multiverse-Specification Diffs}
\label{sec:diff}
 After making changes to a universe during debugging, the analyst may experience difficulty remembering all their universe edits and locating where to place edits in the multiverse specification. We design \diff to support abstracting and propagating edits in the universe to the multiverse specification (DG 3).

\begin{figure*}[t]
  \centering
  \includegraphics[width=\linewidth]{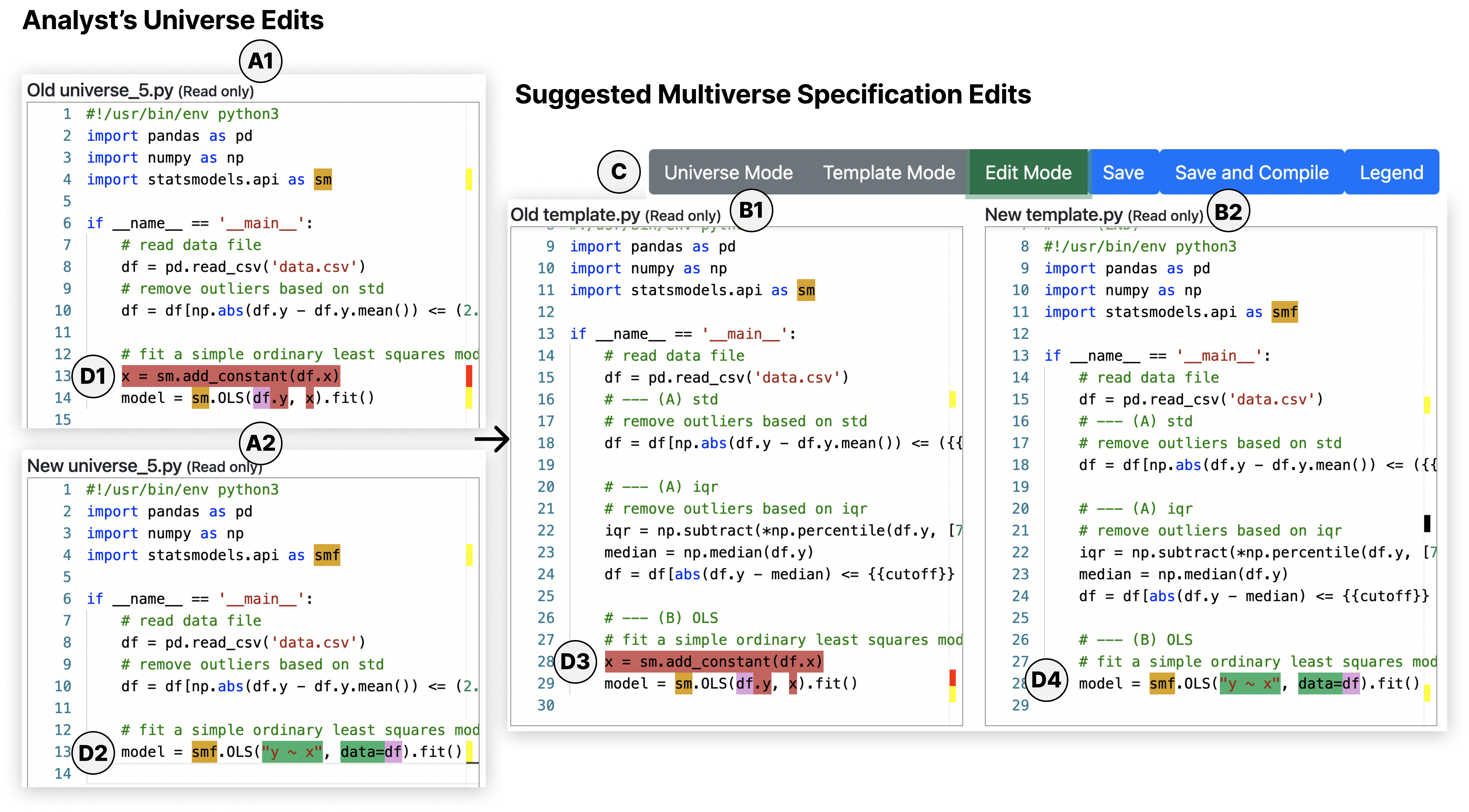}
  \caption{\textbf{\diff Interface}. An analyst makes changes to their universe to use the statsmodels formula API and runs \texttt{boba diff} to abstract and propagate their fixes back to the multiverse specification. This loads the various visual components in this figure. The analyst can view their edit changes via marked up code panels that show the code differences between the old unedited universe (A1) and the new edited universe (A2). Red highlighted code indicates delete edits while green highlighted code indicates insert edits. Yellow highlights show update edits and pink highlights show move edits. The analyst can then navigate via the navigation buttons (C) to view the suggested edits to the new multiverse specification (B2), the contents of which is generated from their universe edits. The interface shows these suggestions by highlighting the edits between the unedited multiverse specification (B1) and the new suggested multiverse specification (B2). Highlights in the old universe matches with those in the old multiverse specification (e.g., D1 and D3). Likewise, highlights in the new edited universe matches with those in the suggested multiverse specification (e.g., D2 and D4). Analysts can make any additional edits to the suggestions in another editor (not shown) before saving the new multiverse specification to disk. Without \diff, analysts would need to remember all their edits in a universe and how those edits propagate to the multiverse specification. \diff makes this process easier by automatically suggesting the necessary propagation of edits.
  }
  \Description{Boba diff figure}
  \label{fig:diff}
\end{figure*}

\diff propagates these edits automatically and presents the changes to the multiverse specification as suggestions. After an analyst finishes making edits to a universe, they can run \texttt{boba diff} to load an interface that communicates the suggested changes (\autoref{fig:diff}). The analyst can then refine these changes further if necessary. 

\diff's interface has three modes. There is a \textit{universe} mode for viewing changes in the universe and a \textit{template} mode for viewing suggested changes in the multiverse specification. The changes are shown as two-panel diffs. Additionally, there is an \textit{edit} mode to make final edits (if necessary) to the suggested changes. The analyst navigates between modes with buttons in the top right (\autoref{fig:diff}C). The analyst may view the \textit{universe} mode to best understand the universe-level changes they made, then proceed to the \textit{template} mode to see how these changes are propagated to the multiverse specification, before finally entering the \textit{edit} mode to finalize suggestions.

In the \textit{universe} mode, the analyst starts with a better grasp of all the edits they made in the universe through viewing a code diff of their universe. The analyst can compare a panel containing highlighted code of the unedited universe (\autoref{fig:diff}A1) with a panel containing highlighted code of the edited universe (\autoref{fig:diff}A2). Highlights to the code show where insertion (green), deletion (red), move (pink), and update (yellow) edits are made (e.g., \autoref{fig:diff}D1-2). 

In the \textit{template} mode, the analyst can view how their changes in the universe are suggested in the multiverse specification. The analyst can compare a panel containing highlighted code of the old multiverse specification (\autoref{fig:diff}B1) with a panel containing highlighted code of the suggested new multiverse specification (\autoref{fig:diff}B2). The highlights for the old multiverse specification are propagated from the unedited universe (e.g., \autoref{fig:diff}D1 to D3). Analogously, the code and highlights for the new multiverse specification are propagated from the edited universe (e.g., \autoref{fig:diff}D2 to D4).

Finally, in the \textit{edit} mode, the analyst can interact with a writable editor that contains the contents of the suggested multiverse specification. We implement a separate mode for editing to encourage a workflow in which the analyst is aware of their changes to the universe and how those changes affect the multiverse specification. To support this further, we include a button for saving the new multiverse specification to disk (based on contents in the editor panel) and a button for saving and compiling only in the \textit{edit} mode.

Beyond navigating these modes, the analyst can navigate between panels via the highlighted code edits. Highlighted code edits that correspond to the same code between panels are linked. For example, move edits in the old and new universe specifications are linked. When a highlighted edit is double-clicked, the linked edit in another panel will appear at the center of editor.

To implement \ourtool, because we need to propagate changes in the universe to specific decision options in the multiverse specification, we must identify decision options in the edited universe. To achieve this, \ourtool compares abstract syntax trees (ASTs) and lines of code of the edited and buggy universe. To compare ASTs, we adapt the gumtree algorithm, a popular source code differencing algorithm based on matching ASTs \cite{falleri_fine-grained_2014-1}\footnote{We release a Python re-implementation of the gumtree algorithm with adaptations for \diff available at \url{https://github.com/behavioral-data/multiverse-tooling/tree/main/src/gumtree}}. To compare lines, we use Python's difflib \cite{difflib} library's mdiff function. Details of the \diff algorithm are in \autoref{appendix:diff}.

\section{Lab Study: Research Questions and Methods}
\label{sec:focused_study}

Using our prototype \ourtool, we conduct a lab study to more specifically understand multiverse debugging workflows.
Our primary goal was not to evaluate \ourtool but rather to create a potential improvement to debugging multiverse analysis in a tangible tool such that analysts could more concretely raise issues, benefits, and design guidelines that are tractable for future tool builders. Additionally, we wanted to allow analysts to explore alternative workflows and elicit responses regarding how features in \ourtool could enable or affect such a workflow.

Three research questions guide our study design and analysis. 
\begin{itemize}
    \item \textbf{RQ1 - Challenges:}~\rqChallengesLong Specifically, do analysts really face the challenges we hypothesized based on prior work, our experiences, and initial correspondences with mutliverse practitioners and tool developers? What additional challenges do they face?
    \item \textbf{RQ2 - Workflows:}~\rqWorkflowsLong
    \item \textbf{RQ3  -  Tool:}~\rqToolAddressChallengesLong~\rqToolLong
\end{itemize}
The first two research questions are more open-ended and exploratory whereas the last research question assesses the benefits of \ourtool's design
interventions and opportunities for further improvement.

\xhdr{Participants} Given that the population of multiverse analysis authors
is relatively small, we focused on recruiting analysts who were interested in
learning about multiverse analysis and represented potential adopters of multiverse analysis. We contacted data analysts through
analysis-related mailing lists at our institution. In the initial interest form, we asked analysts to self-rate their statistical background on a 5-point scale (higher being more familiar). In this scale 4 described analysts who have taken graduate-level courses related to statistical analysis, and 5 described analysts having multiple years of experience with real-word projects involving statistical and data analysis. We also asked analysts to rate their familiarity with R or Python on a 5-point scale with 1 ``being equivalent to have taken an introductory course'' and 5 being having ``5+ years of industry experience''. From the interest forms, we further selected participants with  strong backgrounds in statistical analysis (self-rated 4s and 5s) and comfort with Python or R. A total of 13 analysts participated and their background is summarized in \autoref{tab:participants}.

\begin{table*}
    \small
  \caption{Participant information. Proficiency was self-rated on a 5-point scale with 5 being the highest.}
  \label{tab:participants}
  \begin{tabular}{llllcc}
    \toprule
    ID & Gender & Occupation/Background & Programming Lang. & Lang. Proficiency & Statistics Proficiency\\
    \midrule
    A01 & Female & Researcher in Data Science  & Python & 4 & 4 \\
    A02 & Male & Masters Student in Data Science & Python & 5 & 5\\
    A03 & Female & Masters Student in Industrial Engineering & Python & 4 & 4\\
    A04 & Female & PhD Student in Information Science & Python & 5 & 5 \\
    A05 & Male & PhD Student in Public Policy & R &3 & 5\\
    A06 & Female & PhD Student in Quantitative Ecology & R &3 & 4\\
    A07 & Female & PhD Student in Psychology & R & 5 & 5 \\
    A08 & Female & Data Analyst in Medicine & R & 5 & 5\\
    A09 & Female & Data Scientist & R & 3 & 4\\
    A10 & Female & PhD Student in Applied Mathematics & Python & 4 & 4 \\
    A11 & Male & Data Scientist & R & 5 & 5\\
    A12 & Male & PhD Student in Biostatistics & Python & 4 & 5 \\
    A13 & Male & Professor in Biostatistics & R & 5 & 5 \\
    \bottomrule
  \end{tabular}
\end{table*}

\xhdr{Procedure} The study was conducted in a lab using a designated MacBook Pro
computer on a 27-inch monitor. We allowed participants to use the programming language (i.e., R or
Python) of their
choice and installed what they needed. Analysts primarily used R Studio or Visual Studio Code for their integrated development environment. Before inviting analysts into the lab, we ensured they were familiar with our setup. We wanted to create a debugging environment that was as close to their own experiences.

For materials, we gathered two real-world multiverses from real-world analyses~\cite{jung_female_2014, li_impact_2019} and we created buggy R and Python versions. To introduce realistic bugs, we searched Stack Overflow \cite{stack-overflow} with relevant keywords and online statistics blogs with consolidated lists of errors ~\cite{statology-r, statology-python} to find common bugs encountered during typical statistical analyses. We make the buggy multiverses publicly available and explain the multiverse preparation process in more detail in \autoref{appendix:finding_bugs}.

The study was structured into an initial tutorial phase, followed by two separate debugging task phases that differ in whether the analyst was introduced to \ourtool and was able to use it. We followed this protocol to observe analyst workflows both prior to introducing \ourtool and afterwards.

At the beginning of the study (tutorial phase), we guided analysts through a tutorial that
introduced the multiverse analysis paradigm and how to use Boba. The
tutorial explained how to specify decisions and decision options using Boba syntax.
To ensure analysts understood the concepts behind multiverse analysis and felt
comfortable using Boba, we asked analysts to update a Boba multiverse
specification to add another decision option. 
We also walked analysts through Boba's compile and execute commands.

Next, we asked analysts to debug a realistic multiverse analysis with bugs (Phase 1). In this first part of the study, 
analysts had 25 to 35 minutes to address as many bugs as they
could with the existing Boba tools. Analysts debugged the first multiverse on their own and then completed a
survey about their experience. 

Afterwards, in Phase 2, the first author gave an overview of
\ourtool and how to invoke each command and use the interfaces. Analysts were explicitly told they were free to debug however they wanted. Subsequently, depending on their progress in the 
first portion (i.e., whether they solved the bugs in the first multiverse), analysts were asked to either continue debugging the first multiverse
or debug a second multiverse. More time was spent in the first portion such that analysts can become familiar with Boba and the multiverse paradigm. This also ensured analysts had time to experience challenges specific to multiverse debugging. Finally, analysts completed a survey about their experience using \ourtool. 

We encouraged analysts to talk about their process when they could. If not, they were regularly prompted to speak about their process and describe their thinking. After each debugging session, we also asked open-ended questions with the objective to understand the processes and challenges of multiverse debugging. We gave analysts minimal help beyond pointing out the tools and resources they have available (i.e., the IDE debugging tools, the Internet, and Boba documentation). If analysts were stuck diagnosing and fixing the bug at the single script level (\autoref{fig:debug_workflow}D3) for longer than 15 minutes, we guided analysts by pointing out what the bug is to allow insights along all parts of the workflow.

The study lasted approximately 2
hours. Analysts received a \$50 Amazon gift card as compensation for their
time. This study was determined exempt through the IRB at our institution. We include all lab study materials in our supplemental material.

\noindent \textbf{Qualitative Coding Process.\xspace}~With the exception of one participant (A10) who did not consent to be recorded, we recorded participants' audio and screens. In addition to writing notes of analysts' behaviors while conducting the study, the first author viewed the recordings and transcribed all episodes of interest to the debugging process. To understand common themes that emerged, we used iterative open coding. The themes we observed highlighted analysts’ challenges in debugging multiverse analysis, workflows that analysts gravitated towards, and finally how \ourtool addressed these challenges.

\begin{figure*}[t]
  \centering
  \includegraphics[width=\linewidth]{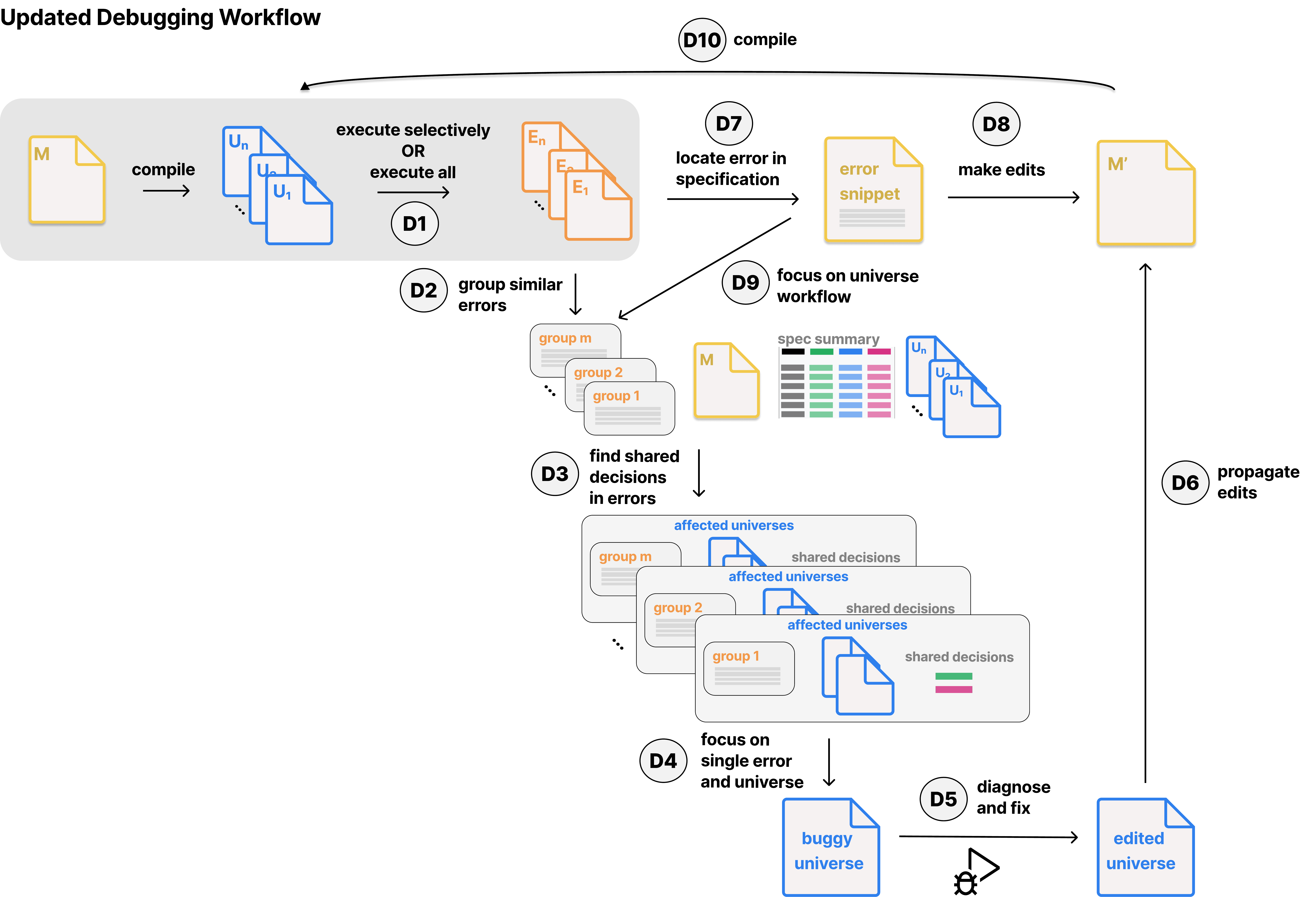}
  \caption{\textbf{Updated Model of Debugging Workflow}. The updated workflow model shows a revised and extended version of the multiverse debugging workflow, enabled through our \rr{lab} study and \ourtool. \rr{Compared to the hypothesized workflow model (\autoref{fig:debug_workflow}), the model derived from the \rr{lab} study has multiple refinements}. First, beyond executing all universes (D1 in \autoref{fig:debug_workflow}), the execution step (D1) now captures analysts' propensity to run a select few universes via \mincover and interest in running their own subset based on specific decision options. Next, our initial understanding of diagnosing the multiverse (D2 in \autoref{fig:debug_workflow}) is expanded to include steps of grouping similar errors (D2), using this grouping along with the specification summary and associated universe information to find shared decisions in error groups (D3), before prioritizing an error and focusing on a single universe (D4). These steps also surfaced an additional challenge of analysts' trouble in understanding the composition of the multiverse. Lastly, to capture analysts' tendency to make fixes directly in the multiverse specification, there is now an additional path in which after observing error messages, an analyst locates the error in the multiverse specification (D7) and then makes the bug patches there directly (D8). Analysts can also go back to the universe workflow (D9) to leverage their comfort with single universe debugging tools. 
  }
  \Description{Updated understanding of workflow}
  \label{fig:updated_workflow}
\end{figure*}

\section{Lab Study: Results}
\label{sec:label_results}
Our lab study identifies four challenges to debugging multiverse analyses and two
approaches analysts take to debug. We also observe how \ourtool affects
analysts' workflows and enables them to overcome the debugging challenges. These
findings inform our updated model of the multiverse debugging workflow, as summarized
in~\autoref{fig:updated_workflow}. The key differences between the updated model and the initial hypothesized model (\autoref{fig:debug_workflow}) are the expanded steps in diagnosing a multiverse (\autoref{fig:updated_workflow}D2-4) (\autoref{subsubsec:result_adress_bugs_workflow}), the additional path to editing a multiverse specification directly (\autoref{fig:updated_workflow}D7-8) (\autoref{subsubsec:use_template}), and the additional choice of selectively executing a semantically meaningful subset of universes (\autoref{fig:updated_workflow}D1) (\autoref{subsubsec:cover_helps}).

\subsection{\rqChallengesLong}

We found that analysts experienced difficulty with two of the three hypothesized challenges: detecting bugs quickly (Challenge 1 in \autoref{subsec:workflow_model_exploratory}) and finding the root causes of bugs (Challenge 2 in \autoref{subsec:workflow_model_exploratory}). In order to find the cause of bugs, we found that analysts needed to group unique errors and identify shared decisions of an error. Maintaining a mental model of the multiverse was also challenging for analysts.

\subsubsection{Minimize latency between executing a multiverse and detecting errors}
\label{subsubsec:results_challenge_latency}
Running the entire multiverse took a non-trivial duration of time, making it
difficult for analysts to receive quick feedback on what errors existed. To
minimize this latency, some analysts picked an arbitrary number of universes to
run [A04, A07, A12]. For instance, prior to using \ourtool, A04 was
reluctant to rerun the multiverse after fixing a bug. Instead, A04 spot-checked three universes. Similarly, A01
reduced the size of the multiverse by commenting out decision options that were
irrelevant to the bug she was addressing.

\subsubsection{Group unique errors and find the number of universes affected}
\label{subsubsec:results_challenge_grp_errors}
In the existing workflow without \ourtool, analysts have no grasp on what the
unique errors are and the number of universes that are affected. Thus,
analysts do not know what a bug fix would even solve and can be left feeling
overwhelmed. A05 captures this perfectly: \shortquote{Seeing that there are 1500 errors but
not having any idea how many were unique makes the process feel overwhelming.} 

Multiple analysts while debugging without \ourtool, and prior to learning about the tool's existence, asked if there was a way to
see the errors grouped together or mentioned lack of grouping as a challenge [A02, A03, A05, A08, A11, A12].

\subsubsection{Identify shared decisions of an error}
Once analysts found an error common across multiple universes, they tried to
isolate the decision choices responsible for producing the error (\autoref{fig:updated_workflow}D3). To do so without \ourtool,
analysts cross-referenced the error messages with the specification summary [A02, A05, A06, A11, A12, A13]. Most participants gave up because the specification summary was ``hard to read'', especially when it contained hundreds of entries with no semantically meaningful structure. 

\subsubsection{Understand the composition of the multiverse}
\label{subsubsec:results_challenge_composition_multiverse}

Understanding the composition of the multiverse means to "understand the components and processes that define and make up this multiverse"~\cite{hall_survey_2022}. For analysts, the composition was not obvious from the information available. To aid in debugging, analysts referenced the multiverse specification file, the specification summary, and the universes to build up a mental map of the multiverse. For A01, this  mental map was essential in her debugging process: \shortquote{Many of these different paths have co-dependencies so I'm not quite sure yet which one of these is truly the issue"}.
To understand common errors in universes, 
analysts consulted error messages and the specification summary to find a common error among several universes. To locate the potential source that caused the error and
understand how a specific universe was generated, analysts looked at the universes, the multiverse specification, and the specification summary. Because the information conveying the composition of the multiverse was scattered, many analysts mentioned processing and navigating
the disjointed information as a challenge [A01, A02, A05, A06, A08, A09, A11, A13]. From just these sources alone, analysts  struggled to construct a mental model of how decision options
were related and contributed to errors common across multiple universes [A01,
A02, A05]. A05 stated how it was \shortquote{not naturally obvious that there
are duplicates stemming from the exact same piece of code}.

As a result, analysts mentioned desiring features that can be broadly categorized into two groups: features that connect information sources and features that can help visualize the multiverse.

For connecting information, analysts desired a feature that
enabled them to locate the code in the multiverse specification which ultimately
resulted in an error [A03, A07, A08]. Similarly, others wanted an
explicit mapping between code in the universe file and code in the multiverse
specification [A02, A03]. For desired visualization features, A11, for example, mentioned wanting a tree structure (like in \autoref{fig:multiverse})
that summarizes the multiverse and associated artifacts: \shortquote{What if I had a tree diagram that I could select which universes does this error happen in that lights up the tree, and show me that they all have this condition.}

\subsection{\rqWorkflowsLong}
\subsubsection{Analysts address bugs in order of error messages but seek new ways to prioritize bugs.}
\label{subsubsec:result_adress_bugs_workflow}

Without \ourtool, analysts often inspected the first error message and set out to fix it [A01, A02, A03, A04, A05, A06, A08, A09, A10, A11, A12, A13]. A01 found this approach comfortable and
reasonable, saying \shortquote{I want to kind of fully tackle that one and then
resolve it and then go on to the next one as opposed to having a higher-level
plan.} However, others wanted a more strategic way to prioritize bugs, which
required a more holistic picture of bugs across multiple universes [A03, A09, A12, A11, A13]. A11 explained his debugging priority was to solve the error affecting many universes: \longquote{I am more interested in spending my time addressing the bugs that occur in several universes versus the bugs in the first universe but I did not have a good sense for how to determine that, so I just went to the first error}

To prioritize, analysts expressed interest in grouping errors together to see unique errors [A02, A03, A05, A08, A11, A12] (\autoref{fig:updated_workflow}D2). Once they have a sense of the unique errors, analysts wanted to see what
was similar and different among universes that encountered the same error in
order to isolate the shared buggy code [A02, A05, A06, A11, A12, A13] (\autoref{fig:updated_workflow}D3). Some analysts [A02, A10, A11] did so by comparing entries in the specification summary that corresponded to universes that had a common error. A02 went so far as to write a script that parsed the specification summary with error ``lines'' (i.e., error messages): 
\longquote{What I was trying to do was to read which (error) lines contain the options and just parse those lines. I was going to write a small script to just parse the lines.}
This idea matches our \errorLogAggregation feature that they had not yet learned about.

\subsubsection{Analysts adopt different strategies based on perceived bug severity.}
\label{subsubsec:use_template}
When analysts perceived an error to have an easy fix, they directly updated the
multiverse specification file without consulting a specific universe script at
all [A02, A03, A04, A07, A10, A13] (\autoref{fig:updated_workflow}D7-8). Analysts stayed in the multiverse specification file because they knew they had to update it eventually anyway. For example, A03 wanted to reduce effort: \shortquote{because the template is the place where we generate the whole universe so I think as long as the bug is fixed in the template, the universe will be free of bugs}. Meanwhile, A07 expressed she preferred
staying in the multiverse specification because she observed a lot of shared code occurred early in the multiverse specification. ~\longquote{I could see that the branching points weren't actually that many if you scroll down through the template file. I saw that there were only really the model points that were breaking routes. If I can get everything before those points to be okay, and then everything subsequently can be re-edited to the template.} When finding errors, analysts also simplified their diagnosing process to just locating the line referenced in the traceback in the multiverse specification (\autoref{fig:updated_workflow}D7). However, because the multiverse specification is not executable, not every bug could be understood and solved there. 

For more involved errors requiring analysts to run large code snippets or inspect intermediate variable values, analysts defaulted to finding and debugging a specific universe. Of the 13 analysts, 12 (everyone except A07) attempted to fix a bug in a specific universe before
making similar fixes to the multiverse specification file. Focusing on one
universe at a time was more familiar to analysts who could rely on their
idiosyncratic debugging approaches, such as using print statements [A02, A03, A04, A12], the interactive debugger [A02, A10], or
the interactive console (i.e., the R console and the Python console) [A03, A05, A06, A08, A09, A11, A13]. Analysts stayed in the same
universe until they fixed a specific bug [A01, A02, A03, A06, A10, A11, A12, A13] or
ensured the universe was completely bug-free [A04, A05, A08, A09].
Once analysts were satisfied with their changes, they updated the multiverse
specification file, re-compiled and re-started the debugging loop.

In some situations, analysts misjudged the complexity of the error and started with the multiverse specification but then went to a universe workflow (\autoref{fig:updated_workflow}D9) after realizing it would have been more effective [A01, A02, A05, A09, A11, A13]. In these cases, analysts wanted to fully leverage their single universe debugging workflows.

\subsection{\rqToolAddressChallengesLong~\rqToolLong}

 Analysts' debugging patterns, which were present without \ourtool but further supported by \ourtool, are described in our updated model of the debugging workflow (\autoref{fig:updated_workflow}). Analysts leveraged \errorLogAggregation to group similar errors (\autoref{fig:updated_workflow}D2), find shared decisions in an error (\autoref{fig:updated_workflow}D3), before then prioritizing an error and focusing on one universe (\autoref{fig:updated_workflow}D4). Moreover, analysts used \mincover to detect errors faster (\autoref{fig:updated_workflow}D1) which inspired them to desire even greater control on what subsets of universes to run. However, analysts seldom used \diff and elected to propagate universe edits manually (\autoref{fig:updated_workflow}D6).

\subsubsection{\mincover reduces latency in detecting bugs and speeds up the development and debugging loop.}
\label{subsubsec:cover_helps}

Nearly all analysts found \mincover feature helpful in expediting the
incremental development and debugging loop [A01, A02, A03, A04, A06, A07, A08,
A09, A10, A11, A12, A13]. Analysts found the \mincover useful for finding the
most common errors quickly and expressed interest in using it as the first step
in debugging multiverse analyses in the future. For example, A07 expressed,
\shortquote{I really like the ability use \texttt{boba ----cover} which helped
pinpoint the most common errors.} Furthermore, for A04, the \mincover enabled
her to work directly in the multiverse specification: \shortquote{These tools
drastically reduced the amount of feedback loop time. Instead of editing the
individual universe files, I mainly worked from the template file.} 

Analysts expressed wanting greater control in specifying which subset of
universes to execute [A01, A04, A07]. Furthermore, other analysts wished they could version
their error messages to maintain the results and errors from a long multiverse run [A01, A12]. 

\subsubsection{\errorLogAggregation helps analysts see unique errors and isolate potential causes to specific decision options}
\label{subsubsec:error_aggr_helps}

Analysts used \errorLogAggregation to identify (i) what the unique errors were
and (ii) how many universes each error message affected. Knowing the unique errors helped
analysts identify familiar error messages they could quickly address [A13] or prioritize error messages that affected
the greatest number of universes [A01, A05, A07, A11]. For instance, A01's strategy was the former:
\shortquote{After seeing the breakdown of the different errors, I would prioritize them and in my head, get a sense of if I fix this fundamental error, would it fix other errors.}

We designed \errorLogAggregation anticipating the challenge of grouping similar errors and finding shared decisions in a common error. 
All 13 analysts liked \errorLogAggregation and said they would want to use it in their workflow.
A05, who was frustrated by his initial lack of awareness of which bugs overlapped with each other,
especially liked the \errorLogAggregation: \shortquote{The error aggregate is
definitely the most useful because it allows for seeing not only the groups of
errors but how many universes are affected.} 

A particularly illustrative example was A02. Prior to using \ourtool, A02 wrote
a custom script to parse the error messages and the specification summary for 15
minutes before running out of time. When he started to use \ourtool, A02 found
\errorLogAggregation especially useful: \shortquote{I really like that you
could get a high-level overview of all the choices that are getting affected.}
Although analysts found \errorLogAggregation beneficial, they also recommended using
visualizations or changing the button layout to make the interface more
intuitive [A01, A03, A06, A12, A13].

\subsubsection{\diff is not as necessary to abstract and propagate patches.}
\label{subsubsec:diff_not_as_helpful}
Analysts found \diff the least useful. One analyst [A02] used the tool to mainly test the feature.  As expected, when analysts stayed in the multiverse specification, \diff was unnecessary. When analysts dove into specific universes, analysts had
mixed feelings about \diff. On one hand, A07, who in her own workflow uses
\texttt{git} diffs only in the CLI, thought \diff would help people who more
``visual.'' On the other hand, A12 thought \diff could be helpful if he
spent more time in a universe and needed to remember more changes: \shortquote{Most of the cases right now you give me are simple but once the debug time is too long then you'll easily forget how you did the changes. That would be the most useful case.} 

\section{Discussion}
\label{sec:discussion}
In this work, we built a prototype tool and conducted a subsequent lab study to understand and address multiverse debugging challenges. From our lab study, which leveraged our tool as a design probe, we developed an updated model of multiverse debugging workflows (\autoref{fig:updated_workflow}). In this section, we synthesize the results from our lab study and share implications for improving multiverse analysis tools. We highlight four key design implications that would better support multiverse debugging, review the limitations of our work, and discuss future work.

\subsection{Design Implications}
\label{subsec:design_implications}

\subsubsection{Tools should reduce the latency in encountering multiverse errors}

The long time to detect an error message (step D1 in \autoref{fig:updated_workflow}) was a challenge we hypothesized (\autoref{subsec:workflow_model_exploratory}) and later confirmed in our lab study (\autoref{subsubsec:results_challenge_latency}). In the lab study, we even found analysts trying their own ways to increase the speed of detecting error messages (i.e., commenting out code). We also found the \mincover feature to be especially useful because it enabled this faster detection (\autoref{subsubsec:cover_helps}). Future tools should consider features that reduce the latency to detect erroneous multiverse code whether that is through something like \mincover or letting analysts run subsets of universes (something we discuss as another design implication in \autoref{subsubsec:find_decisions_universes})

\subsubsection{Tools should summarize unique errors and highlight shared decision options}
The challenge of understanding what unique errors exist (step D2 in \autoref{fig:updated_workflow}) and what are common decision options (step D3 in \autoref{fig:updated_workflow}) was pervasive in the lab study (\autoref{subsubsec:results_challenge_grp_errors}). As a result, \ourtool’s \errorLogAggregation feature which directly addresses this was appreciated by all analysts (\autoref{subsubsec:error_aggr_helps}). Multiverse debugging tools will benefit from some form of error message aggregation.

\subsubsection{Tools should help analysts understand the composition of the multiverse}

A key challenge that surfaced among analysts in the lab study was understanding the composition of the multiverse; that is, how the specification of decisions and options led to the
generation of universes (\autoref{subsubsec:results_challenge_composition_multiverse}). While we hypothesized the need to understand the multiverse would contribute to the cognitive load in propagating edits (\autoref{subsec:workflow_model_exploratory}), our lab study revealed this understanding is critical much earlier in the debugging cycle (\autoref{fig:updated_workflow}D2-4) and less important when propagating edits (\autoref{fig:updated_workflow}D6) (\autoref{subsubsec:diff_not_as_helpful}). Specifically, in diagnosing an error message, analysts needed this comprehension to begin understanding what decision options may have caused an error or how code may be shared across certain universes as a result of the multiverse specification. Moreover, multiple analysts expressed connecting the multiverse structure (i.e., showing the structure relating universes, decisions, and decision options) to the multiverse specification code as something that would aid in their debugging process (\autoref{subsubsec:results_challenge_composition_multiverse}).

Informed by our lab study, future tools that aid in understanding the composition of the multiverse should connect the multiverse structure with the multiverse specification. One opportunity to support understanding is through interactive visualizations that connect a visualization of the multiverse structure 
with a visual representation of the multiverse specification code. Such a visualization would also support analysts' iterative authoring process \cite{kery_variolite_2017, kery_interactions_2018}, enabling analysts to understand how the composition of the multiverse changes over time as a result of code changes. Prior work has also highlighted the need for real-time and interactive visualization of the multiverse structure \cite{sarma_multiverse_2021}.

While researchers have started to develop multiverse-specific visualizations \cite{liu_boba_2021, hall_survey_2022}, none have focused on interactions showing the multiverse structure and the specific code implementing them in the multiverse specification. Future work should explore how to best communicate the specified multiverse structure in relation to the specification code.

\subsubsection{Support Analysts in Finding Relevant Universes and Decision Options in the Multiverse} 
\label{subsubsec:find_decisions_universes}

Another common theme observed in the lab study was analysts’ need to have control in finding subsets of universes or subsets of decision options. For example, to better isolate a potential cause for an error message, analysts expressed wanting to know what subset of universes to run that correspond to specific combinations of analysis decisions (\autoref{subsubsec:cover_helps}). This is difficult because to find that subset, analysts currently need to either consult the specification summary and navigate through hundreds of entries or write custom functions to parse this information. On the other hand, \ourtool’s \errorLogAggregation feature, which analysts ubiquitously found helpful (\autoref{subsubsec:error_aggr_helps}), is a realization of finding a subset of meaningful decision options from a subset of universes. 

Therefore, core activities involved in multiverse debugging require finding a subset of universes based on specified decision options or finding a subset of decision options based on specifying a subset of universes. Tools that enable this process would improve analysts' capability to and speed in diagnosing error messages. As such, future tools should incorporate effective multiverse selection based on universe or decision option constraints.

\subsection{Limitations}

\ourtool focuses on extending Boba to understand multiverse debugging workflows. Therefore, its features are all command-line based. For analysts who are less comfortable with programming and more comfortable with workflows that involve graphical user interfaces (e.g., Stata \cite{stata}, SPSS \cite{spss}), \ourtool may be difficult to use. 

We note several limitations of our user study. First, the study had a small sample size and consisted of people new to multiverse analysis. As the number of people who perform multiverse analysis is small, we determined an in-person lab study
was the best way to gather people, provide a tutorial on multiverse analysis and get them up to speed with existing tools. Results, therefore, might be different for multiverse experts. However, as multiverse analysis is a relatively new analysis paradigm, there are very few experts to date and an important focus lies on empowering a broad set of analysts to employ multiverse analyses. Multiverse analysis is targeted to those familiar with statistical practices who may want to adopt this paradigm (which is our lab study population) and it is through making the associated challenges easier (specification, analyzing results, and debugging) that this paradigm will receive greater adoption. Prior tools \cite{liu_boba_2021, sarma_multiverse_2021, rdfanalysis} improved workflows surrounding specification and analyzing results but that adoption is still limited in part due to debugging challenges that are not yet supported \cite{sarma_multiverse_2021}. Understanding the debugging challenges of a potential adopter is one step toward this larger goal.

Additionally, in order to facilitate a lab study of reasonable duration, we chose to conduct a same-day in-person study of 2 hours and give analysts a largely pre-written multiverse. Future work should explore debugging processes based on a multiverse the participant is developing themselves as well as more complex multiverses. Finally, while the bugs introduced into the pre-written multiverses reflected common analysis errors, they may not be representative of those encountered in more complex or domain-specific analyses. We hypothesize that the overall workflow will likely be similar but analysts may want to focus even more on debugging individual universes. In addition, \diff may be more useful in these larger multiverses with more complex bug fixes.

\subsection{Future Work}
\label{sec:future_work}
\xhdr{Towards enabling debugging for larger classes of bugs}
\ourtool helps analysts author a multiverse that is free from execution errors. 
However, there could be bugs that do not lead to execution errors, including bugs around statistical analysis misspecification (e.g., a poorly specified model and model formula). These bugs may not raise error messages but threaten the statistical validity of the analysis. This type of bugs is not specific to multiverse analysis but relevant to all analysis paradigms. Recent tools have been developed to improve statistical validity in traditional analysis~\cite{jun_tea_2019, jun_tisane_2022} but more work is needed to help analysts detect such bugs. Another class of bugs is related to errors in multiverse specification. For example, an analyst may have intended to perform data filtering only for a subset of models but did not specify that constraint in the multiverse specification. While there would not be any execution errors, the universes affected may not reflect the intended analysis. Future work could explore how to detect and communicate these bugs to the analyst.

\xhdr{Exploring the trade-offs between universe level and multiverse level workflows} 
While most analysts favored debugging with a single universe, we discovered in our lab study some analysts tended to debug with the multiverse specification directly (\autoref{subsubsec:use_template}). Analysts' tendency to focus on one level could also be influenced by the tool they are working with.
Boba \cite{liu_boba_2021} naturally encourages a universe level workflow as the universes are separated from the multiverse specification and are no different than traditional analysis scripts. This lets analysts use their favorite tools and familiar workflows. However, the separation has the drawback that the multiverse specification cannot be directly executed. \mverselib \cite{sarma_multiverse_2021}, in contrast, encourages a multiverse level workflow and lets analysts run universes via library functions in the same file in which the multiverse is specified. However, placing everything in one file puts multiverse specification logic and analysis code all in a single file, which may even more difficult to debug. Future work should explore these trade-offs between executable higher-level multiverse specifications and the complexity of navigation and debugging.

\section{Conclusion}

This paper focuses on debugging as a key, under-scrutinized barrier to broader multiverse analysis adoption. To understand analysts' challenges and debugging workflows, we build a prototype debugging tool, \ourtool, and conduct a qualitative lab study using \ourtool as a probe. This work contributes the first user study to better understand, model, and support the unique challenges that multiverse analysis poses for debugging. In addition, we provide an open-source tool, \ourtool, that alleviates some of the observed challenges.  We synthesize findings to develop a model of multiverse debugging workflows and associated challenges (\autoref{fig:updated_workflow}) and highlight design implications for future tools to support multiverse analysis debugging.

\begin{acks}
We are grateful to our anonymous reviewers for their thoughtful comments. We would also like to thank the members of the UW Behavioral Data Science group and the UW PLSE group for their feedback on this work. This research was supported in part by NSF IIS-1901386, NSF CAREER IIS-2142794, NSF CNS-2025022, NIH R01MH125179, Bill \& Melinda Gates Foundation (INV-004841), the Office of Naval Research (\#N00014-21-1-2154), a Microsoft AI for Accessibility grant and a Garvey Institute Innovation grant.
\end{acks}

\bibliographystyle{ACM-Reference-Format}
\bibliography{references_custom}

\input
\clearpage
\appendix

\section{Initial Correspondences with Multiverse Experts}
\label{appendix:formative}
\subsection{Interviews with Two Multiverse Practitioners}
To identify specific challenges in authoring and conducting multiverse analyses, we first conduct in-depth interview studies with two researchers who have recently authored multiverse analyses. We found these researchers through our collaboration networks. Neither relied on existing multiverse tools. Instead,
they wrote custom scripts that generated each universe script. During the interviews, which lasted for
approximately two hours each, the researchers walked us through their analyses,
including their scripts, findings, and any historical artifacts from their git repository 
histories. 
Without being prompted, 
both brought up how challenging finding and propagating bug fixes is for them.

We learned that the researchers approach authoring multiverse analyses in a
bottom-up, iterative fashion. They focus on a few key decisions and
options, consult their peers and supervisors, and then add additional
decisions and options based on their team's input. This iterative nature
requires keeping track of which combinations of decision options were
previously considered and how, if at all, the results have altered since
changing or adding decisions and decision options. The same process applies when the
researchers encounter and fix bugs. They must identify bugs, fix decision
options that introduce the bugs, and then re-run their multiverse analyses
to see how the bugs impact their results.

This led to an understanding that multiverse debugging is a key challenge and that resolving difficulties surrounding this process could make it easier to author multiverse analyses more generally.
\subsection{Additional Correspondences with Experienced Multiverse Tool Developers}
We cross-examined our observed challenges and insights in debugging with two independent, experienced
researchers who have authored multiverse analyses and developed multiverse
analysis tools.  We corresponded with
these researchers via email.

Both researchers corroborated the importance of
starting with a single universe and then propagating changes to the rest of the
universes: \shortquote{I may
look at a single universe. Then I apply the solution to all affected paths.
Currently, this can only be achieved by modifying the multiverse
specification.} The other researcher had a similar debugging process: \shortquote{I
always debug by looking at individual universe scripts that instantiate a
particular set of decisions that I think might be involved in the error}.
They also mentioned how debugging multiverse analyses is like debugging a
single universe but with \shortquote{the added difficulty of figuring out why
the bugs come up in a particular analysis}. Finally, one tool developer also highlighted the additional steps needed to pinpoint an error: \shortquote{I often read the error messages and pick a specific error to focus on. Then I examine all paths that lead to a specific error to distill commonality.}

\section{\mincover Algorithm}
\label{appendix:min_cover}
\begin{algorithm}
\caption{Decision Cover}\label{alg:min_cover}
\SetAlgoLined
\SetKw{KwInit}{Initialize}

\KwIn{${S} = \{u_i \mid i=1\dots n\}$ (set of universes in the entire multiverse), ${D} = \{d_j \mid j=1\dots\ m\}$ (set of decision options in the multiverse)
        \tcp*[r]{Each universe $u_i$ is represented by a unique set of decision options $D_{u_i} \subset D$ }}
\KwInit{$M \gets \emptyset$}

\While{$S \neq \emptyset$}{
 $u \sim Uniform(S)$ 
 $M \gets M \cup \{u\}$
 $T \gets \{u\}$
\For{$v \in S$} {
    \If{$D_{u} \cap D_{v} \neq \emptyset$}{
         $T \gets T \cup \{v\}$
    }
}
$S \gets S \setminus T$
}

\Return{$M$}
\end{algorithm}

The \mincover algorithm is an iterative loop of sampling a universe from the multiverse and reducing the multiverse by removing all universes that contain decision options of universes sampled so far. Algorithm \autoref{alg:min_cover} summarizes the
\mincover algorithm. We start with the set of all universes. Until this set is empty, a universe is randomly sampled and all universes that share any decision option will be removed from this set. We take the set of sampled universes as the reduced set of universes to run.

\section{Algorithm for Universe-to-Multiverse-Specification Diffs}
\label{appendix:diff}
\subsection{Boba Background}
There are two main ways to specify decisions in the \textit{template} file: \textit{placeholder variables} for decision options that
can be placed in-line and \textit{code blocks} for decision options that involve multiple lines of code. Placeholder variables can
be placed anywhere in the template file. Users specify the placeholder decision name and its alternative options. During
compilation, Boba removes the placeholder identifier and replaces it with one of its alternative values. In \autoref{boba_authoring}A, the
\texttt{cutoff} and \texttt{brm\_family} decisions are defined with placeholder variables. Meanwhile, a decision block is used to specify
multiple versions of a code block that act as alternative decision options for one analytical decision. For example, in
\autoref{boba_authoring}A, the \texttt{Model} decision block consists of two alternative code blocks, representing an option for a \textit{frequentist}
model and an option for a \textit{bayesian} model. When compiling the \textit{template} file, Boba will instantiate only one code
block corresponding to a decision in a universe.

\subsection{Algorithm}

\ourtool compares abstract syntax trees (ASTs)
and lines of code of the edited and unedited universe. ASTs provide the granularity
needed to identify decision options and potential changes to these options that are specified in-line (Boba placeholder
variables) in the new universe. Meanwhile, comparing code at the line granularity helps locate decision options specified by multiple
lines of code (Boba code blocks). Furthermore, comparing lines also helps map universe code blocks to the multiverse
specification blocks. 

 We use information from the compilation process to know where in the unedited old universe the Boba variables are located and the split points between Boba code blocks. In short, we have a mapping between the unedited universe and the multiverse specification. We then find where in the new edited universe the locations of Boba variables are via AST matching and locations of Boba code blocks via line matching. Through the mapping between old and new universe, we can then map changes in the new universe all the way back to the multiverse specification.

To pinpoint code changes in the universe that correspond to decision options specified inline in the multiverse specification, we match the ASTs of the unedited and edited universes. Matching ASTs  provides additional granularity than line difference algorithms and enables direct mappings between code that corresponds to matched subtrees in the AST. We use gumtree~\cite{falleri_fine-grained_2014}to find code in the new universe that corresponds to Boba variables. If changes exist, these are mapped to the multiverse specification.

We use the Python difflib \cite{difflib} library's mdiff function to match the
start of code blocks between the old and new universe files. For each line in the old universe if it is matched with the new universe and it is the start of the block boundary, we add the new universe line as the start of the corresponding Boba block. If the line is deleted and it is at the boundary of the Boba block, we add the next line in the new universe. Finally, if a new line is inserted and it is at the start of a new block, we always default to including it at the start of a new block. With our initial multiverse specification and unedited universe mapping, we can propagate edits in the universe back to the multiverse specification.

The \diff algorithm based on gumtree’s AST matching algorithm is best suited for small to medium edit changes. As these edits are common in most of bug fixes, gumtree is an adequate choice. 

\section{Process for Finding Bugs for the Lab Study}
\label{appendix:finding_bugs}
We gathered two multiverses from which we created buggy R and Python versions. The first multiverse, \textsc{Hurricane}, is authored by Simonsohn \textit{et al.} \cite{simonsohn_specification_2020}
and challenges the reported analysis in a previous study~\cite{jung_female_2014}. The study explored whether hurricanes with female names resulted in more deaths. The second multiverse, \textsc{Reading}, is an example from Boba \cite{liu_boba_2021}. \textsc{Reading} is based on how researchers of a published paper~\cite{li_impact_2019}, on whether different web layouts result in faster reading speeds, might construct a multiverse from their analysis.

To introduce realistic bugs, we first identified common bugs encountered during
typical statistical analyses. We searched Stack Overflow \cite{stack-overflow} to find errors. For R, we searched Stack Overflow with tags \texttt{R} and keyword \textit{error} to find relevant posts. Similarly, for Python, we searched with tags \texttt{Python}, \texttt{pandas}\cite{pandas}, and \texttt{statsmodels}\cite{statsmodels} and the keyword \textit{error} to find relevant posts. In addition to Stack Overflow, we consulted an online statistics blog with consolidated lists of Python~\cite{statology-python} and R errors~\cite{statology-r}. 

This resulted in errors that encompass data parsing, data splitting, and model
specification. The R version of \textsc{Hurricane} included 5 errors. One was a syntax error, one was a logical one-off error, two more errors were errors that resulted from poor data processing, and the last error was a model fit error due to a poorly specified model formula. The Python version contained 3 errors: the same one-off error, a data processing error, and the same model fit error. 

For the \textsc{Reading} multiverse, the R version involved 3 errors: two errors related to poor data/model specification, and a third error with misspecified data transformation. The Python version had 3 errors as well: an error as a result of using the wrong model, an error with the wrong syntax for data filtering, and a third error from parsing the data improperly. We include all lab study materials in our supplemental material.

\end{document}